\documentstyle[preprint,floats,epsf,aps,prb]{revtex}
\begin{document}

\draft


\title {\bf CO on Pt(111) puzzle; A possible solution}
\author{Ilya Grinberg, Yashar Yourdshahyan, and Andrew M. Rappe}
\address{Department of Chemistry and Laboratory for Research on the
Structure of Matter\\ University of Pennsylvania, Philadelphia, PA
19104--6323} 
\date{\today} 
\maketitle

\begin{abstract}

CO adsorption on the Pt(111) surface is studied using first-principles
methods.  As found in a recent study {[Feibelman {\em et al}.,
J. Phys. Chem. B {\bf 105}, 4018 (2001)]}, we find the preferred
adsorption site within density functional theory to be the hollow
site, whereas experimentally it is found that the top site is
preferred.  The influence of pseudopotential and exchange-correlation
functional error on the CO binding energy and site preference was
carefully investigated. We also compare the site preference energy of CO on Pt(111) with the 
reaction energy of formaldehyde formation from H$_2$ and CO.
 We show that the discrepancies between the
experimental and theoretical results are due to the generalized
gradient approximation (GGA) treating different bond orders with
varying accuracy.  As a result, GGA results will contain significant
error whenever bonds of different bond order are broken and formed.


\end{abstract} 


\section{Introduction}

The interactions of small molecules on metal surfaces play an
important role in many industrial processes, such as automotive
catalysis, corrosion, tribology, and gas sensing.  These processes
have been the subject of intensive experimental and theoretical
investigation~\cite{SomorjaiBook}.  Due to its simplicity, carbon
monoxide frequently serves as a probe molecule in studies aiming to
understand the nature of molecule-metal interactions.  The insights
gained from CO/metal surface studies can then be transferred to more
complex systems. A particularly well-studied system is CO on the
Pt(111) surface.  The adsorption of CO on Pt(111) has been studied
experimentally at various coverages and
pressures.~\cite{DiNardo,SomorjaiSFG} Scanning tunneling microscopy
(STM), low energy electron diffraction (LEED), and infrared adsorption
(IR) experiments have shown that in ultra-high vacuum conditions CO
adsorbs on top sites exclusively at low coverages and forms a
$c(4\times 2)$ top-bridge overlayer at half-monolayer
coverage~\cite{Ogletree,Blackman,Ryberg,Hopster,Steininger,Hayden,Luo,Xu,Henderson,Bocquet,Stroscio}.
Yeo {\em et al}.\ measured the CO chemisorption energy calorimetrically,
finding that it decreases from about 1.9~eV at low coverages to 1.2~eV
at half-monolayer coverage~\cite{Yeo}.

Several phenomenological models of CO-metal surface bonding have been
proposed.  The most widely used conceptual framework for small
molecule adsorption on surfaces is the Blyholder
model~\cite{Blyholder}, with electron donation from the CO 5$\sigma$
orbital to the metal and back donation from the metal to the CO
2$\pi$*, coupling the CO levels to the metal $sp$ states and $d$
states.  Building on the Blyholder model, Hammer {\em et al}.\ have
proposed a quantitative model for CO chemisorption energy in terms of
the energy of the center of the metal $d$ bands~\cite{Hammer}.  Their
model gives reasonably good agreement with experiment.  However, the
Blyholder model has been recently challenged by Fohlisch {\em et al}.,
who have studied CO adsorbed on the top, bridge and hollow sites on
Ni(100) with X-ray emission spectroscopy and {\em ab initio}
calculations~\cite{Fohlisch}.  They conclude that chemisorption energy
is the result of a balance between the repulsive $\sigma$ and
attractive $\pi$ interactions, leading to very different electronic
structures for different adsorption sites, despite similar binding
energies.

Density functional theory~\cite{HK,KS} (DFT) calculations have also
been carried out to study this system at both the local density
approximation (LDA)~\cite{CA,PZ} and generalized gradient
approximation (GGA)~\cite{PBE,RPBE} levels.  Jennison {\em et al}.\
studied CO adsorption on a 91-atom Pt cluster simulating a (111)
surface~\cite{Jennison}.  They found the top site to be preferred,
followed by bridge, with hcp less stable.  Philipsen {\em et al}.\
found a 0.24~eV difference between the binding energies of the top and
hcp sites, with the top preferred when relativistic effects were
included in the calculation~\cite{Philipsen}.  Even though their study
was done using periodic boundary conditions, they only used two metal
layers to represent the platinum surface in their calculations.  Lynch
and Hu carried out a GGA study of the adsorption of CO on various
Pt(111) surface sites~\cite{Hu}. In this study, the bottom two layers
of Pt were frozen at bulk positions, but the top layer was allowed to
relax.  Using the same four-site, three-layer approximation they
obtained chemisorption energies for CO on fcc, hcp, top and bridge
sites at quarter monolayer coverage.  They obtained a top site binding
energy of 1.89~eV, with bridge $E_{\rm chem}$ of 2.00~eV, hcp of
2.00~eV and 2.09~eV for the fcc site.  The preference of hollow sites
over the top site as well as the values of the binding energies
are not in agreement with experiment.

In a recent study, Feibelman {\em et al}.\ showed~\cite{Feibelman}
that the results of Jennison {\em et al}.\ and Lynch {\em et al}.\ as
well as other earlier calculations suffered from several convergence
problems.  When approximations such as basis set size, $k$-point
sampling and slab thickness were converged, DFT calculations on the
CO/Pt(111) system using a variety of functionals in pseudopotential
(PSP), projector-augmented wave (PAW) and full-potential linearized
augmented plane wave (FLAPW) approaches consistently preferred the
hollow site over the top site.  Their results are summarized in Tables
I and II.  While the hollow site was always preferred, the magnitude
of the energy difference varied depending on the core electron
approximation and exchange-correlation functional used.  LDA
calculations give about 0.4~eV for the site preference energy ($E_{\rm
t-h}$), while GGA results vary from 0.25~eV obtained with PW91/PBE
ultra-soft potential calculations, to 0.13~eV obtained with PW91/PAW
to 0.10~eV obtained with FLAPW calculations.  Feibelman {\em et al}.\
also report calculations for CO/Cu(111) and CO/Rh(111), where GGA
calculations again prefer the fcc site, while experimentally the top
site is preferred.  GGA and experiment do agree in the case of
Ru(0001), where GGA predicts the experimentally-observed adsorption on
the top site. In the case of CO/Pd(111), GGA predicts the
experimentally-observed fcc site adsorption.  They conclude that
GGA/LDA tends to favor higher coordination, which is correct in the
case of Pd(111) but incorrect for surfaces where the top site is
preferred. However, the reasons for discrepancies between LDA and GGA
results and between PSP, PAW, and FLAPW calculations are not resolved,
nor do they propose a way to estimate the site preference error.  The
PSP-GGA DFT method has proven to be reliable and accurate for a wide
variety of solid-state systems, so the inability of GGA calculations
to predict site preference of CO on metal surfaces is puzzling.

\begin{table}[t]
\caption{Results of DFT calculations of Feibelman {\em et al}.~\cite{Feibelman} for
site preference energy for various coverages and methods. $\Delta
E_{\rm t-h}$ is the CO/Pt(111) difference between top and hcp site
chemisorption energies.  All energies are in eV.}

\begin{tabular}{crrrr}
 &\multicolumn{1}{c}{$\Theta$(ML)}&\multicolumn{1}{c}{method}&\multicolumn{1}{c}{XC}&\multicolumn{1}{c}{$E_{\rm t-h}$}\\
\hline

$3\times 2\sqrt 3$&  1/12& VASP,USP&    PW91&  0.25\\

$2\times 2$&           1/4&  DACAPO,USP&  PW91&  0.23\\

$2\times 2$&           1/4&  DACAPO,USP&  PBE&   0.24\\

$2\times 2$&           1/4&  DACAPO,USP&  RPBE&  0.16\\

$2\times 2$&           1/4&  DACAPO,USP&  LDA&   0.45\\

$c(4\times 2)$&        1/4&  VASP,USP&    LDA&   0.41\\

$c(4\times 2)$&        1/4&  DACAPO,USP&  PW91&  0.23\\

$c(4\times 2)$&        1/4&  VASP,USP&    PW91&  0.18\\

$c(4\times 2)$&        1/4&  VASP,PAW&    PW91&  0.13\\

$\sqrt 3 \times \sqrt 3$&  1/3&  DACAPO,USP&    PW91&  0.23\\

$\sqrt 3 \times \sqrt 3$&  1/3&  WIEN,FLAPW&    PW91&  0.10\\
\end{tabular}
\end{table}

\begin{table}[t]
\caption{Calculated normal mode frequencies (cm$^{-1}$) for CO on
Pt(111) at top, bridge and fcc sites, from calculations of Feibelman
{\em et al}. ~\cite{Feibelman} Experimental results for the top and the bridge sites are taken from Ref. 5 and Ref. 6, and for the hollow site from Ref. 28}

\begin{tabular}{ccccc}
& \multicolumn{1}{c}{$\omega_{\rm C-O}$ Exp.}
& \multicolumn{1}{c}{$\omega_{\rm C-O}$ DFT}
& \multicolumn{1}{c}{$\omega_{\rm C-Pt}$ Exp.}
& \multicolumn{1}{c}{$\omega_{\rm C-Pt}$ DFT} \\
\hline
top&          2095&  2140&  466&   461\\

bridge&       1845&  1926&  370&   393\\

fcc&          1750&  1832&    &    342\\
\end{tabular}
\end{table}

The results from the pioneering work of Feibelman {\em et al}.\ raise
several questions.  First, what is the reason for the discrepancies
between PSP and all-electron results, and can a closer agreement
between PSP and all-electron calculations be obtained? Second, are the
site preference errors limited to CO on (111) surfaces, or are similar
effects found with other molecules and/or other surfaces?  Third, does
the error in $E_{\rm t-h}$ come from an error in $E_{\rm chem}$ of a
particular site, or a combination of errors in $E_{\rm chem}$ of both
sites? The answers to all of these questions will help identify the reason for DFT
failure for CO/Pt(111).

In PSP-DFT calculations there are only two uncontrolled
approximations: the replacement of the nucleus and the core electrons
with a pseudopotential, and the form of the exchange-correlation
functional. The effect of pseudopotential error on adsorption energies
has not been examined but has been assumed to be small.
Exchange-correlation functional error has been known to affect binding
energies, most notably for LDA.  Hammer {\em et al}.\ studied the
effect of different exchange-correlation functionals on the binding
energy of CO on the fcc site of the Pd(111) surface~\cite{RPBE}. They
found that LDA overbinds by about 1.0~eV, while PBE GGA overbinds by
0.2--0.3~eV and RPBE and revPBE GGA overbind by 0.1--0.2~eV.  The source
of the GGA and LDA DFT site preference errors must lie in either one
or both of the uncontrolled approximations. For FLAPW calculations
there is no pseudopotential error, so the wrong site preference must
come entirely from the functional error.  For PSP calculations, both
PSP and XC functional errors can contribute to error in chemisorption
energy, with the PSP and XC effects either cooperating or canceling
out. In the next section, we examine PSP error.

\section{Pseudopotential Effects}

Pseudopotential error can affect different bonds in the CO/Pt(111)
system differently.  In addition to freezing the core states, the
pseudopotential approach involves changing the nuclear potential
inside a core radius $r_c$.  If the wave function of one atom interacts
with the potential inside the core region of another atom, the
unphysical nature of the pseudopotential will be manifested in the
results.  The CO bond will be affected by overlap of core radii for
typical carbon and oxygen pseudopotentials, due to the very short CO
bond length (1.12--1.19~\AA  ).  However, the C--Pt and Pt--Pt bonds
will not have core overlap, since these bonds are longer,
1.85--2.1~\AA  and 2.77~\AA  respectively.  Thus, core overlap in the
CO bond introduces an additional, albeit controlled error into PSP-DFT
calculations which can either cooperate or cancel out with the XC
functional error.  Calculations by Feibelman {\em et al}.\ suggest
that for $E_{\rm t-h}$ DFT error is dominant, while PSP error affects
$E_{\rm t-h}$ by 0.15~eV at most.

To investigate the effects of PSP core overlap, we carry out a series
of calculations on the CO/Pt(111) system.  We use several carbon and
oxygen pseudopotentials, gradually eliminating core overlap from the
CO bond, while keeping the same Pt pseudopotential for all
calculations.  This allows us to obtain the accuracy limit of DFT
calculations using the pseudopotential approach.

We carry out DFT calculations on the CO/Pt(111) system at quarter
coverage using optimized norm-conserving pseudopotentials~\cite{RRKJ}
and the PBE GGA functional~\cite{PBE}. The platinum pseudopotential
was created using a wave function averaging relativistic
construction~\cite{GRRrel}, with the designed non-local
method~\cite{DNL} used to achieve good norm and eigenvalue
transferability~\cite{GRRnorm}.  We use an 81~Ry plane-wave cutoff for
PSP~1 through PSP~5.  The unusually high plane-wave cutoff allows us
to use carbon and oxygen pseudopotentials without core overlap in CO.
All carbon and oxygen pseudopotentials were created from the same
reference configuration, differing in their core radii; PSP~6 and
PSP~7 were created at 50~Ry cutoff.  We use five metal layers with six
layers of vacuum to model the Pt(111) surface, keeping the three
bottom layers fixed and relaxing the coordinates of the top two
layers. All calculations are done in a $c(4\times 2)$ unit cell using
a $4\times 4 \times 1$ grid of Monkhorst-Pack $k$-points~\cite{MP} to
sample the reducible Brillouin zone.  For every set of carbon and
oxygen PSPs, we calculate the binding energy at the top and hollow
sites.  The differences $E_{\rm t-h}$ are shown in Figure~1;
additional properties are listed in Table~III.  The positive values of
$E_{\rm t-h}$ mean that the hcp site is lower in energy than the top
site.

\begin{table}[t]
\caption{Results of DFT-GGA calculations for the CO/Pt(111) system for
various carbon and oxygen pseudopotentials. PSPs 1-5 were created with 
81 Ry plane-wave cutoff, and PSPs 6 and 7 were created with 50 Ry plane-wave cutoff.  Shown are the CO chemisorption 
energy ($E_{\rm chem}$), CO bond length ($R_{\rm CO}$), 
and the site preference energy $E_{\rm
t-h}$. The values for the hollow site are shown in
parentheses. Energies are in eV, core radii are in $a_0$, and bond lengths
are in \AA .}  

\begin{tabular}{crrrr}
& \multicolumn{1}{c}{$r_c^{\rm O}$,$r_c^{\rm C}$} 
& \multicolumn{1}{c}{$E_{\rm chem}$}
& \multicolumn{1}{c}{$R_{\rm CO}$} 
& \multicolumn{1}{c}{$E_{\rm t-h}$} \\ 

& \multicolumn{1}{c}{($a_0$)}
& \multicolumn{1}{c}{(eV)}
& \multicolumn{1}{c}{(\AA)}
& \multicolumn{1}{c}{(eV)} \\ \hline

PSP 1&        0.50,0.58&  1.73 (1.81)&   1.165 (1.202)&    0.082\\

PSP 2&        0.58,0.64&   1.73 (1.81)&   1.164 (1.201)&   0.079\\

PSP 3&        0.64,0.69&   1.73 (1.80)&  1.163 (1.200)&   0.074\\

PSP 4&        0.69,0.74&   1.73 (1.80)&   1.161 (1.196)&  0.068\\

PSP 5&        0.74,0.79&   1.73 (1.78)&   1.154 (1.194)&  0.055\\

PSP 6&        0.74,0.79&   1.72 (1.75)&   1.148 (1.181)&  0.030 \\

PSP 7&        0.79,0.85&   1.72 (1.73)&   1.146 (1.172)&   0.007\\

AE&		     &               &               &  0.100\cite{Feibelman}\\     

Exp&		     &	 \multicolumn{1}{c}{1.68$\pm$0.12\cite{Yeo}}
\end{tabular}
\end{table}

As can be seen from Figure~1, the hcp site is always preferred by
0.01--0.09~eV, in disagreement with experiment but in agreement with
calculations by Feibelman {\em et al}.  As expected, the effect of
core overlap is exponential in $r_c$.  This implies that past a
threshhold of 1.25--1.35 $a_0$ (r$_c$$^O$ + r$_c$$^C$) a small change in $r_c$ will produce a
significant error in calculated DFT properties. It is gratifying to
note that $E_{\rm t-h}$ converges as we eliminate core overlap and
that the $E_{\rm t-h}$ obtained by us using the best carbon and oxygen
pseudopotentials is essentially identical with that by Feibelman {\em
et al}.\ using FLAPW calculations. We have also used different Pt PSPs
with the same C and O PSPs, but found only a small (about 0.02~eV)
variation of $E_{\rm t-h}$ with Pt PSP. Our results suggest that PSP
calculations with no core overlap preserve almost all of the accuracy
of an all-electron approach, with a PSP error of less than 0.02~eV.
However, using pseudopotentials unconverged with respect to core
overlap can lead to significantly larger PSP error (0.1~eV or more) in
$E_{\rm t-h}$ as well as in other properties.

\section{Generality of site preference error}   

A second question raised by the results of Feibelman {\em et al}.\ is
the generality of site preference errors. It is obviously important to
know if the failure of DFT/GGA is confined to CO on (111) metal
surfaces, or if this effect is found in other types of systems. We
first examine the results in the literature for CO on (100) metal
surfaces.  Eichler and Hafner found in PSP-GGA calculations of CO
adsorption on Rh(100)~\cite{Eichler} that at quarter monolayer
coverage the bridge and the hollow sites were preferred over the top
site, while experiment showed the top and bridge site to be occupied
at low coverages with no hollow site occupation. The site preference
energies from their calculations are summarized in Table~IV.  They
conclude that it is possible that {\em ab initio} calculations
overestimate the energetic preference for bridge and hollow site
adsorption and underestimate the height of the barrier for the
migration from the top site to the bridge site.

For Cu(100) experimentally only top site adsorption is seen.  Recently
Favot {\em et al}.\ examined the adsorption of CO on Cu(100) using LDA
and GGA~\cite{Favot}. The site preference energies from their
calculations are summarized in Table~IV.  They find $E_{\rm t-h}$ of
0.18~eV for LDA and $E_{\rm t-h}$ of $-0.09$~eV for GGA approximations,
showing the same trend of a reduction in $E_{\rm t-h}$ on going from
LDA to GGA as the results of Feibelman {\em et al}.  It therefore
seems very likely that the explanation for the discrepancies between
top-hollow site preference in CO/(100) metal surface systems and the
top-hollow site preference in CO/(111) systems are of the same origin.

\begin{table}[t]
\caption{Results of DFT calculations for CO adsorption on (100) metal surfaces.}

\begin{tabular}{crrrr}
& \multicolumn{1}{c}{$E_{\rm chem}^{\rm top}$}
& \multicolumn{1}{c}{$E_{\rm chem}^{\rm bridge}$}
& \multicolumn{1}{c}{$E_{\rm chem}^{\rm hollow}$}
& \multicolumn{1}{c}{$E_{\rm t-h}$}\\
& \multicolumn{1}{c}{(eV)}
& \multicolumn{1}{c}{(eV)}
& \multicolumn{1}{c}{(eV)}
& \multicolumn{1}{c}{(eV)}\\
\hline

CO/Cu(100) LDA~\cite{Favot}&      	 1.1&  1.15&  1.28& 0.18\\

CO/Cu(100) PBE~\cite{Favot}&          0.78& 0.75&  0.69& -0.09\\

CO/Rh(100) PW91 ~\cite{Eichler}&        2.0&  2.15& 2.2&  0.20\\
\end{tabular}
\end{table}

To explore the generality of this error, it is instructive to see how
DFT performs for a gas phase system which exhibits a similar bond
breaking/formation pattern.  We therefore examine the energy of
formaldehyde formation ($\Delta E_{\rm form}$) from CO and H$_2$ ($\rm
H_2 + CO \rightarrow H_2CO$).  To remove the pseudopotential error, we
gradually reduce the core radii of the carbon and oxygen
pseudopotentials to eliminate core overlap, while keeping the same
hydrogen PSP.  The results from these calculations are presented in
Figure~1 and Table~V. Positive values of E$_{\rm form}$ indicate
that formaldehyde is lower in energy than H$_2$ and CO.

\begin{table}[t]

\caption{Results of DFT-GGA calculations for the reaction $\rm H_2 +
CO \rightarrow H_2CO$ reaction for various carbon and oxygen
pseudopotentials. PSPs 1-5 were created with 81 Ry plane-wave cutoff, and PSPs 6,7 were created with 50 Ry plane-wave cutoff.  Shown are the CO bond energy ($BE_{\rm CO}$), CO
bond length ($R_{\rm CO}$) and the energy of the reaction ($E_{\rm form}$). The CO  bond length in formaldehyde is shown in parenthesis. All experimental results taken from the CRC Handbook\cite{CRC}, unless otherwise noted. Energies are in eV, core radii are in $a_0$ and bond lengths are in \AA .} 

\begin{tabular}{crrrr}
& \multicolumn{1}{c}{$r_c^{\rm O}$,$r_c^{\rm C}$} 
& \multicolumn{1}{c}{$BE_{\rm CO}$}
& \multicolumn{1}{c}{$R_{\rm CO}$} 
& \multicolumn{1}{c}{$E_{\rm form}$} \\ 

& \multicolumn{1}{c}{($a_0$)}
& \multicolumn{1}{c}{(eV)}
& \multicolumn{1}{c}{(\AA)}
& \multicolumn{1}{c}{(eV)} \\ \hline

PSP 1&        0.50,0.58&  11.758&   1.150(1.221)&   0.546\\

PSP 2&        0.58,0.64&  11.835&   1.149(1.220)&   0.550\\

PSP 3&        0.64,0.69&  11.836&   1.147(1.219)&   0.538\\

PSP 4&        0.69,0.74&  11.881&   1.144(1.217)&   0.528\\

PSP 5&        0.74,0.79&  11.967&   1.142(1.213)&   0.512\\

PSP 6&        0.74,0.79&  12.200&   1.142(1.201)&   0.512\\

PSP 7&        0.79,0.85&  12.409&   1.122(1.193)&   0.424\\

AE   &               &11.660\cite{Kurth} &\multicolumn{1}{c}{1.136\cite{Fohlisch2}}(1.209)\cite{Deng}&      0.46\cite{Deng}  \\

Exp  &	             &   11.27&   1.128(1.210)&     0.30\cite{Deng}\\

\end{tabular}
\end{table}

Figure~1 demonstrates that $\Delta E_{\rm form}$ depends exponentially
on core overlap, similarly to $E_{\rm t-h}$.  Increasing the core
radii lowers both $\Delta E_{\rm form}$ and $E_{\rm t-h}$, although
the range of $\Delta E_{\rm form}$ values is larger than the range of
$E_{\rm t-h}$ values. We find a converged $\Delta E_{\rm form}$ of
0.55~eV, considerably overestimating the experimental $\Delta E_{\rm
form}$ of 0.30~eV.  Thus we see that DFT/GGA calculations fail for a
variety of calculations involving the CO molecule.
	
\section{Accuracy of the chemisorption energy at the top and hcp sites} 

Since $E_{\rm t-h}$ is the difference of two binding energies, error
in $E^{\rm chem}_{\rm top}$ and/or $E^{\rm chem}_{\rm hollow}$ can
lead to a wrong $E_{\rm t-h}$.  We now consider how the chemisorption
error varies with site.  For quarter monolayer coverage, our converged
$E_{\rm chem}^{\rm top}$ is 1.729~eV.  This agrees well with the
experimental adsorption energy of $1.68\pm 0.12$~eV obtained by Yeo
{\em et al}.~\cite{Yeo} for the CO/Pt(111) system at $\Theta=0.25$,
where CO molecules are adsorbed mostly on the top
sites~\cite{Steininger}, with only a small bridge site population.
The hcp binding energy at quarter coverage is not known, but it must
be smaller than the adsorption energy at the top or bridge sites.  The
experimental difference between $E_{\rm chem}^{\rm top}$ and $E_{\rm
chem}^{\rm bridge}$ is estimated to be about
0.06eV~\cite{DiNardo,Mieher,Schweizer}.  Since the bridge site is
populated at low coverages, while the hcp site is not, $E_{\rm
chem}^{\rm hcp}$ must be smaller than $E_{\rm chem}^{\rm
bridge}$.Therefore, $E_{\rm chem}^{\rm hcp}$ is at most $1.63\pm
0.12$~eV.  Our converged value for $E_{\rm chem}^{\rm hcp}$ is
1.811~eV. We see that the $E_{\rm chem}^{\rm top}$ is overestimated by
about 0.05~eV, while $E_{\rm chem}^{\rm hcp}$ is overestimated by at
least 0.18~eV.  Since the $E_{\rm chem}^{\rm top}$ is within the
experimental error bars, we conclude the wrong $E_{\rm t-h}$ is due
primarily due to the erroneous value $E_{\rm chem}^{\rm hcp}$.  This
demonstrates that for CO on Pt(111), DFT-GGA methods give significant
error for adsorption on the hcp site, but the top site is treated
accurately.  We will now address the cause of DFT-GGA failure.

\section{Bonding Contributions to Chemisorption Site Preference}

Since DFT-GGA is inaccurate in a similar way for a variety of CO/metal
systems and an organic reaction, we investigate whether there is a
common reason for this failure.  To understand the causes of DFT-GGA
failure, first consider the nature of $E_{\rm t-h}$.

The site preference energy $E_{\rm t-h}$ is the difference in the
chemisorption energies of the CO molecule on the two surface sites.

\begin{eqnarray}
 E_{\rm t-h} =  E_{\rm top}  -E_{\rm hollow}
\end{eqnarray}

$E_{\rm t-h}$ is also the energy of CO migration from the top site to
the hollow site, in which the CO bond is weakened, C--Pt bonds
are formed and the platinum surface atoms rearrange to accommodate the
CO.  We can therefore write the $\Delta E$ of the reaction as the sum
of energies of bonds broken and bonds formed.
 
\begin{eqnarray}
E_{\rm t-h}= \Delta E^{\rm CO}_{\rm t-h} + \Delta E^{\rm C-Pt}_{\rm
t-h} + \Delta E^{\rm Pt-Pt}_{\rm t-h} ,
\end{eqnarray}

\noindent
where $\Delta E^{\rm CO}_{\rm t-h}$ is the difference in CO bond
energy when adsorbed on top and hcp sites, $\Delta E^{\rm C-Pt}_{\rm
t-h}$ is the difference between bond energies of the lone C--Pt bond
on the top site and the three C--Pt bonds on the hollow site, and
$\Delta E^{\rm Pt-Pt}_{\rm t-h}$ is the energy cost of rearrangement
of the Pt--Pt bonds.

To correctly predict the site preference energy, DFT calculations must
either treat all bonds in Eq.~2 accurately or have the errors in
various terms fortuitously cancel out.  The CO/Pt(111) system is very
challenging, since metals exhibit diffuse metallic bonding while the
CO molecule is an example of very tight, covalent bonding.  The
accuracy of LDA and GGA calculations is known to diminish as the
electronic charge density becomes more inhomogeneous, so the CO,
metal-metal and metal-carbon bond energies will not be estimated with
the same accuracy.  Once the nearly perfect error cancellation is
lost, a significant error in one bond energy that is not matched by an
error in another bond energy will lead to a wrong $E_{\rm t-h}$.
Similarly, the formaldehyde formation reaction involves changing the
CO bond from a triple bond to a double bond, breaking of the H$_2$
bond, and forming two C--H bonds.  In order to compute the formation
energy, these bonding changes must all be modeled accurately or with
errors that cancel.

\section{Bond-order changes and DFT-GGA Accuracy}

In practice GGA does not perform equally well for the energetics of
all these bonds.  DFT-GGA calculations are almost always very good for
geometry optimization, due to the fact that the inhomogeneity of the
electron gas does not significantly change as bonds shorten or
lengthen slightly.  Thus, the errors due to the use of an approximate
functional cancel out.  However, in calculating the energy difference
of structures with different inhomogeneities, the error cancellation
will not work as well~\cite{Mitas}.  In chemical language, this change
in the character of the inhomogeneous electron gas between reactants
and products is known as bond order change.

The effect of bond order change on the accuracy of DFT energies can be
seen from the work of Kurth {\em et al}., who recently examined the
performance of various exchange-correlation functionals, LDA, GGA and
meta-GGA~\cite{PKZB} on atomization energies of small
molecules~\cite{Kurth}.  They performed all-electron calculations, so
deviations from experiment in the values of the atomization energies
are due to functional error only. We show their data for LDA, PBE,
RPBE and PKZB functionals in Table~VI.  Since we want to evaluate the
quality of DFT $\Delta E$ results for reactions with molecular
reactants and products, we are more interested in the accuracy of
atomization energy (or bond energy) differences, than in the errors in
atomization energies themselves.

From the data of Kurth {\em et al}., it is clear that XC functionals
perform well on bond energy differences with similar bond order,
e.g. a C--H bond and an N--H bond.  The energy differences between
bonds of different bond order, e.g CO and NO, are considerably less
accurate.  To illustrate this, we plot the PBE error in bond energy in
Figure~2.  Inevitably, a few bond energies have very similar error
differences, such as CO - C$=$C. However, all the energy differences
between bonds of different bond order involving oxygen have large
errors.  The CO - NO energy difference error is 0.413~eV, and the CO -
O$_2$ error is 0.596~eV, twice as large as the CO - N$_2$ error of
0.226~eV, and almost an order of magnitude larger than the CH - OH
error of 0.0510~eV.  The high quality of single bond energy results
and the decrease in bond energy error from the double bond region to
the triple bond region in Figure~2 implies that PBE calculations are
accurate for first-bond energies, significantly overestimate the
second-bond energies, and underestimate the energy of third bonds.
Therefore, significant bond order changes are accompanied by large DFT
errors.  The inaccuracy of DFT in computing the relative energies of
systems with different bond orders has also been noted by Mitas in his
work on silicon clusters.~\cite{Mitas}

\section{The cause of DFT-GGA Failure}

We can now examine the source of E$_{\rm t-h}$ discrepancy in the
CO/Pt(111) system.  In Eq.~2, three terms contribute to $E_{\rm t-h}$:
$\Delta E^{\rm CO}_{\rm t-h}$, $\Delta E^{\rm C-Pt}_{\rm t-h}$ and
$\Delta E^{\rm Pt-Pt}_{\rm t-h}$.  The Pt--Pt bond order changes
little from top site chemisorption to hcp site, and DFT is known to do
well on metallic systems, so we rule out these bonds as a source of
$E_{\rm t-h}$ discrepancy.  The calculations of Feibelman {\em et
al}.\ (Table~II) show a C--Pt frequency of 461~cm$^{-1}$ for the top
site, 393~cm$^{-1}$ for the bridge site, and 342~cm$^{-1}$ for the
hollow site.  The DFT results compare very well with experimental
values of 466~cm$^{-1}$ for the top site and 385~cm$^{-1}$ for the
bridge site.  Since the top-bridge C-Pt bond order change is similar
to the bridge-hcp C-Pt bond order change, and both the top and the
bridge C-Pt DFT frequencies are of high quality, this suggests that
the hollow site frequency is likely to be accurate as well.  The
calculated DFT C--Pt distances are accurate within 0.02~\AA  for both
sites.  The high quality of frequency and bond length results implies
that the C--Pt bond energies at both sites are accurate as well.
Thus, there is at most a small error in the $\Delta E^{\rm C-Pt}_{\rm
t-h}$.

Turning to the CO bond, the experimental CO stretch frequencies
(Table~II) show significant chemical changes due to migration from the
top to the hollow site. The CO bond at a top site is only slightly
weaker than the gas-phase CO triple bond, while the CO bond on the
hollow site is closer to a gas-phase CO double bond than to a triple
bond.  The free CO molecule has a vibrational frequency of
2140~cm$^{-1}$, the HCO radical CO stretch frequency is
1865~cm$^{-1}$, and a typical double bond CO stretch has a frequency
of about 1700~cm$^{-1}$ in organic molecules. The experimental top
site CO stretch frequency in the CO/Pt(111) system is 2095~cm$^{-1}$,
the bridge site CO stretch frequency is 1845~cm$^{-1}$, and the hollow
site frequency can be estimated at around
1770~cm$^{-1}$~\cite{Weaver}.  Molecular orbital analysis assigns the
gas-phase CO a bond order ($BO$) of 3, HCO $BO=2.5$ and C$=$O $BO=2$.
Interpolation based on vibrational frequencies allows us to estimate
the top site CO $BO=2.9$ and hollow site CO $BO=2.2$. This implies a
rather significant CO bond order change of 0.7.  Since, as shown by
the results of Kurth {\em et al}., PBE underestimates the energy of
the third CO bond, the energy cost of CO bond order change from 2.9 to
2.2 will be underestimated.  This will lead to an overestimation of
the $E_{\rm chem}^{\rm hcp}$ and consequently to the incorrect $E_{\rm
t-h}$. In the same fashion, the underestimation of the energy cost of
the third CO bond breaking will lead to an overestimation of $E_{\rm
form}$.

\begin{table}[t]
\caption{Results of all-electron DFT calculations of Kurth {\em et
al}.~\cite{Kurth} \ for atomization energies of small molecules with various
exchange-correlation functionals.  All energies are in eV.}

\begin{tabular}{crrrrr}
 &\multicolumn{1}{c}{Exp.}&\multicolumn{1}{c}{LDA}&\multicolumn{1}{c}{PBE}&\multicolumn{1}{c}{RPBE}&\multicolumn{1}{c}{PKZB} \\
&\multicolumn{1}{c}{}&\multicolumn{1}{c}{Error}&\multicolumn{1}{c}{Error}&\multicolumn{1}{c}{Error}&\multicolumn{1}{c}{Error} \\
\hline

H$_2$&      	 4.76&   0.16&  -0.21&    -0.17&  0.22\\

CH$_4$&         18.23&   1.87&  -0.03&   -0.39&  0.08\\

NH$_3$&         12.93&  1.74&   0.19&  -0.18&   0.06\\

OH &             4.63&   1.77&   0.14&   -0.01&    0.06\\

H$_2$O&         10.10& 1.49&  0.08&   -0.25&   -0.10\\

HF&              6.12&  0.93&    0.05&   -0.14&    -0.09\\

CO&		11.27& 1.73&   0.42&  -0.06&   -0.14\\

N$_2$&           9.94& 1.69&   0.63&  0.18&    0.03\\

NO&		 6.65&  1.99&   0.82&   0.38&     0.24\\

O$_2$&           5.24&  2.37&   1.01&   0.56&    0.47\\

C$_2$H$_2$&	17.63& 2.38&  0.38&  -0.22&    -0.19\\

C$_2$H$_4$&     24.46& 3.06&  0.39&  -0.35&    -0.05\\
\end{tabular}
\end{table}

A comparison of our converged top and hollow site chemisorption
energies supports the assignment of DFT failure to the CO bond.  As
shown above, the $E^{\rm chem}_{\rm top}$ value is accurate within the
experimental error bars, while $E^{\rm chem}_{\rm hcp}$ is off by at
least 0.18~eV.  The CO bond is only slightly weakened upon adsorption
in the top site, as shown by experimental red-shift of only about
50~cm$^{-1}$ from the gas-phase CO stretch frequency and a bond order
change of 0.1 from the free CO.  Therefore, the accurate value for
$E_{\rm chem}^{\rm top}$ can be considered the result of very good
DFT-GGA error cancellation in CO bond energetics.  For the hcp site,
with a red-shift of about 390~cm$^{-1}$ and CO bond order change from
3.0 to 2.2, the inaccurate chemisorption energy coincides with more
dissimilar CO bonds.  

The identification of $\Delta E^{\rm CO}_{\rm t-h}$ underestimation as
the major cause of site preference error is also supported by the
vibrational frequency results of Feibelman {\em et al}.  The DFT
calculations show a shift of 214~cm$^{-1}$ in CO bond stretch
frequency from top to bridge site, and a shift of 308~cm$^{-1}$ from
top to hcp site.  These results are lower than the experimental
250~cm$^{-1}$ shift from top to bridge and the 325~cm$^{-1}$
shift from top to hcp site.  Since stronger bonds will usually have
higher vibrational frequencies, this underestimate of the frequency
shift provides additional evidence that GGA underestimates the energy
cost of the third CO bond breaking.

\section{Conclusion}

We have shown that for molecule-surface systems, converged PSP
calculations and all-electron calculations yield very similar results.
The frozen core creates only a minimal inherent error, but core radius
overlap can lead to larger deviations. An examination of the
literature shows that DFT energy errors are not limited to CO
adsorption on (111) metal surfaces, but occur for many reactions
involving CO.  We have also shown that the underestimation of CO bond
energy loss in migration from the top to the hollow site is largely
responsible for the incorrect site preference obtained by DFT-GGA
calculations.  The dependence of DFT error on bond order change
implies that the top site chemisorption energy obtained by DFT-GGA
calculations is accurate, while the chemisorption energy of the hollow
site is not and leads to incorrect $E_{\rm t-h}$. This is confirmed by
our converged PSP-DFT results for E$_{\rm chem}^{\rm top}$ and E$_{\rm
chem}^{\rm hcp}$. Since $\Delta E_{\rm triple-double}^{\rm CO}$ is
independent of metal surface, errors in this energy will affect the
$E_{\rm t-h}$ site preference energies on many metal surfaces.  We 
therefore propose that a simple empirical correction,  based on reaction
energies of small organic molecules,  may permit accurate prediction of site preference.


Similar errors will also be found for other small molecules which
experience changes in molecular bond order.  The PKZB meta-GGA functional~\cite{PKZB} may predict the
right site preference without empirical corrections, due to its
superior performance in calculating the atomization energies of small
molecules.


\section{Acknowledgments}
This work was supported by NSF grant DMR 97--02514 and the Air Force
Office of Scientific Research, Air Force Materiel Command, USAF, under
grant number F49620--00--1--0170.  AMR acknowledges the support of the
Camille and Henry Dreyfus Foundation.  Computational support was
provided by the San Diego Supercomputer Center, National Center for
Supercomputing Applications and NAVOCEANO MSRC.

\newpage
\begin{figure}[p]
\epsfysize=3.0in
\centerline{\epsfbox[0 0 402 357]{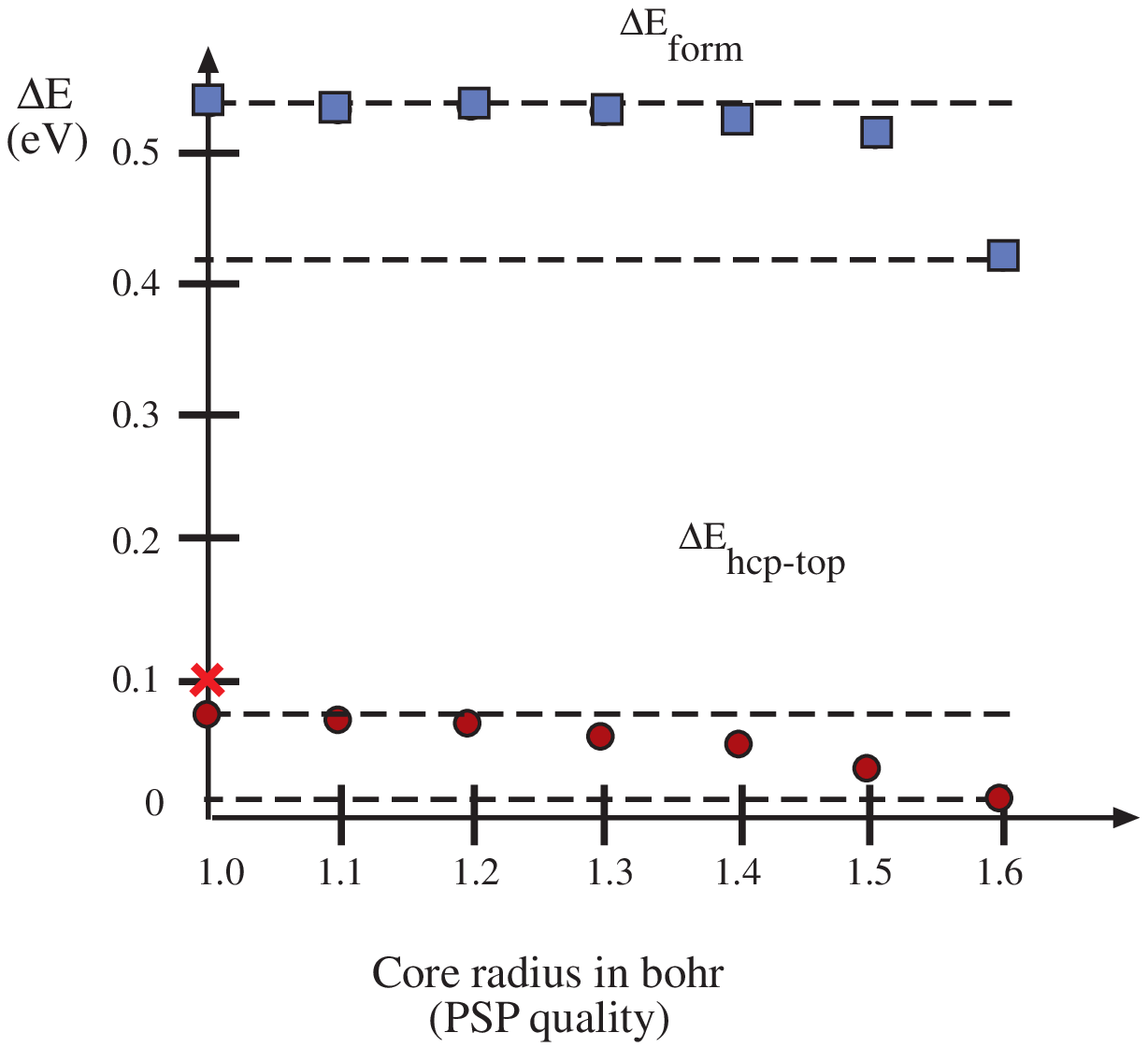}}
\caption{H$_2$CO and CO/Pt(111) PSP convergence.
Circles correspond to site preference $\Delta$E. Squares correspond to $\Delta$E of formaldehyde formation. $\boldmath \times$ marks the Feibelman {\em et al}.  LAPW result for site preference $\Delta$E.}

\end{figure}

\newpage
\begin{figure}[p]
\epsfysize=3.0in
\centerline{\epsfbox[0 0 593 631]{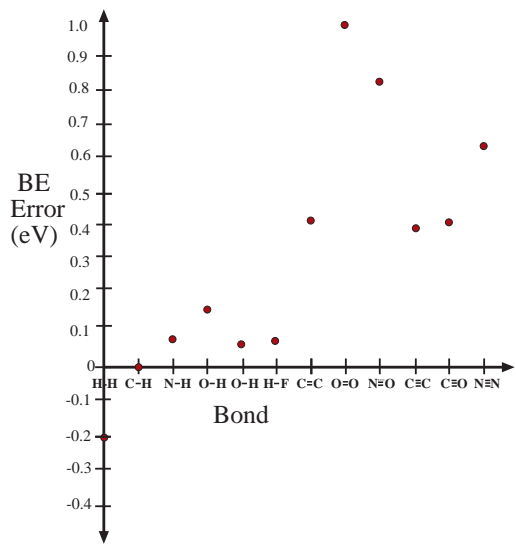}}
\caption{PBE error in bond energies from calculations on small molecules.}

\end{figure}

%

\end{document}